\documentclass[aps, prl, superscriptaddress, showpacs, reprint]{revtex4-1}
\usepackage{float}
\usepackage{color, graphicx}
\usepackage{amssymb,amsmath, amsfonts, bm}
\usepackage[colorlinks,citecolor=blue, urlcolor=blue, linkcolor=blue]{hyperref}

\begin{document}

\title{Magnetic-Order Induced Spectral-Weight Redistribution in a
  Triangular Surface System}

\author{Gang Li}
\affiliation{\mbox{Institut f\"ur Theoretische Physik und Astrophysik, 
  Universit\"at W\"urzburg, 97074 W\"urzburg, Germany} }

\author{Philipp H\"opfner}
\author{J\"org Sch\"afer}
\author{Christian Blumenstein}
\author{Sebastian Meyer}
\affiliation{\mbox{Physikalisches Institut, Universit\"at W\"urzburg, 97074
  W\"urzburg, Germany}}

\author{Aaron Bostwick}
\author{Eli Rotenberg}
\affiliation{\mbox{Advanced Light Source, Lawrence Berkeley National
  Laboratory, Berkeley, California 94720, USA}}

\author{Ralph Claessen}
\affiliation{\mbox{Physikalisches Institut, Universit\"at W\"urzburg, 97074
  W\"urzburg, Germany}}

\author{Werner Hanke}
\affiliation{\mbox{Institut f\"ur Theoretische Physik und Astrophysik, 
  Universit\"at W\"urzburg, 97074 W\"urzburg, Germany} }

\begin{abstract}
The Sn-induced $\sqrt3 \times \sqrt3$ surface reconstruction on
Si(111) has been investigated by material-specific many-body
calculations and by angle-resolved photoelectron spectroscopy
(ARPES). This triangular surface system in the low adatom coverage
regime is governed by rather localized dangling bond orbitals with
enhanced electronic correlations and it is prone to exhibit magnetic
frustration. We find a rather good overall agreement of the spectral
function and its temperature-dependence between theory and
experiment. Changes in the ARPES band topology in comparison to the
density functional calculations can be explained as a spectral weight
redistribution with respect to an additional symmetry which is not
due to any geometrical change. This new symmetry corresponds to
a magnetic order, which is found to be more complex than the canonical
$120^{\circ}$ anti-ferromagnetic order on a triangular lattice with
nearest-neighbor coupling only. 
\end{abstract}

\maketitle

The adatom-covered surfaces of Si(111) and Ge(111) provide an
excellent playground to study the competition and cooperation of
geometrical frustration and electronic correlations in quasi-two
dimensional triangular systems. The deposition of $1/3$ of a monolayer
of the group IV elements Pb and Sn renders these surfaces $\sqrt3
\times \sqrt3$ reconstructed \cite{1996Natur.381..398C,
  PhysRevLett.79.2859, PhysRevB.62.8082}. These adatoms induce states
in the semiconducting band gap, including surface bands which are well
separated from the bulk bands.  These states are dominated by highly
localized dangling bond orbitals, and their narrow band nature leads to
enhanced electronic correlations in these systems.  
Moreover, long-range magnetic ordering is generally suppressed by
magnetic frustration inherently contained in such effectively
triangular surface lattices \cite{PhysRevLett.98.086401}. 

In the $\alpha$-phase of Pb/Ge(111) \cite{1996Natur.381..398C} and Sn/Ge(111)
 \cite{PhysRevLett.79.2859}, a transition from a $\sqrt3
 \times \sqrt3$ to a $3 \times 3$ structure occurs upon cooling, which can
be explained within a dynamical fluctuation model
\cite{PhysRevLett.82.442}. Pb or Sn atoms oscillate vertically at room
temperature (RT), but their motions are frozen out at low temperature
(LT), leaving two out of three Pb (Sn) atoms per unit cell at a different
height than the third one.  
A similar structural transition might thus be expected to be observed on
the surface of Sn/Si(111). 
However, Sn/Si(111) shows no $3 \times 3$ signature in low-energy
electron diffraction (LEED) and scanning tunneling microscopy (STM)
experiments at both RT and LT, down to 6\,K \cite{PhysRevB.62.8082,
  PhysRevB.65.201308}.  
Careful analysis of photoelectron diffraction results further 
revealed that all Sn adatoms have the same bonding geometry
\cite{PhysRevLett.98.126401}. 
  
In addition,  in Sn/Si(111) the observed valence-band
photoemission spectra show a shadow band at
$\overline{\Gamma}_{\sqrt{3}}$ (see Fig. \ref{EDC_Comp} for
notations) and an approximate $3\times3$ periodicity of the overall
surface band topology \cite{PhysRevB.62.8082,PhysRevB.68.235332}.
Also, experimentally a clear conductance dip at the Fermi level was observed in
scanning tunneling spectroscopy at low temperature \cite{PhysRevLett.98.126401}. The ground
state, thus, is believed to be a narrow gap insulator.
Calculations based on density functional theory (DFT) in the local-spin density
approximation (LSDA) failed to explain both facts, i.e. a
\emph{metallic} ground state without backfolded band features at
$\overline{\Gamma}_{\sqrt{3}}$ is predicted
\cite{PhysRevB.59.1891,PhysRevB.62.1556}. 
The insulating ground state can be explained by the strong electronic
correlations favored by the small bandwidth of the Sn/Si(111) surface
band. In LSDA + U \cite{PhysRevLett.98.086401} 
and more sophisticated local density approximation (LDA) + many-body calculations based on the
Hubbard model \cite{PhysRevB.82.035116, PhysRevB.83.041104}, the
ground state is confirmed to be insulating and identified as a narrow
gap \emph{Mott insulator}.
However, the shadow band feature and the additional band symmetry
observed in ARPES have not been explained consistently with
respect to all the above mentioned experimental facts. 
In particular, the inherent spin frustrations contained in Sn/Si(111)
further elaborate the issue \cite{Morita2002, PhysRevLett.60.2531,
  PhysRevB.83.041104} with a remaining lack of knowledge on their
interplay with electronic correlations.

In this Letter, we provide a consistent explanation for the surface
band topology and the shadow band to originate from electronic
correlations rather than from structural aspects. We provide direct
evidence from our temperature-dependent calculations that the
approximate $3\times3$ symmetry observed in ARPES is indeed a
consequence of magnetic correlations.  
However, the order associated with it is more complicated than the
standard $120\,^{\circ}$ anti-ferromagnetic (AFM) spin-arrangement for
triangular surfaces.

\begin{figure}[htbp]
\centering
\includegraphics[width=\linewidth]{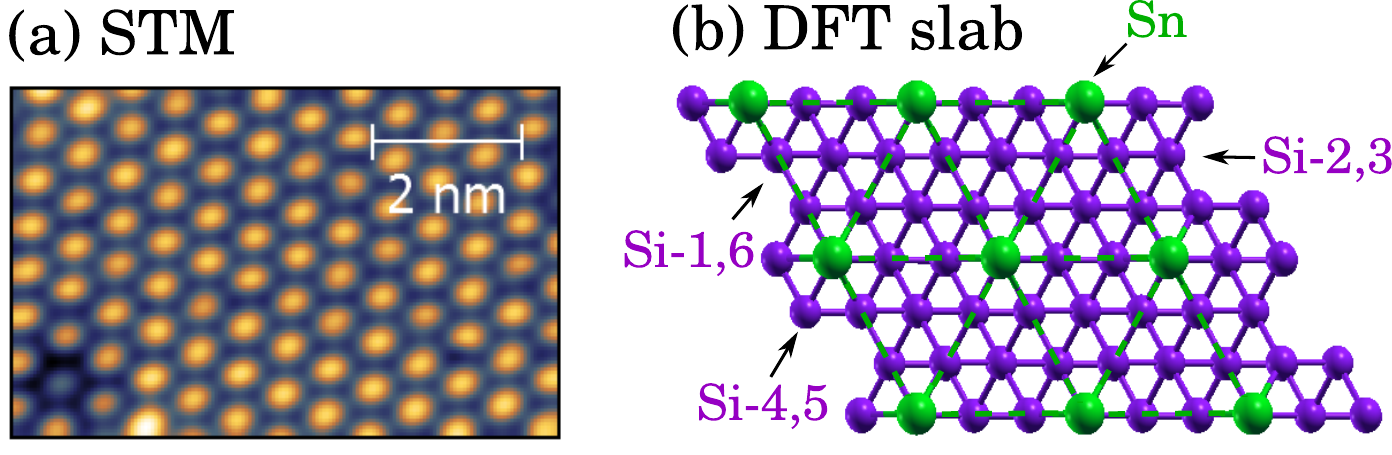}
\caption{ 
(a). STM image of the Sn/Si(111) surface at $T = 300$\,K
    (bias: -1.0\,V; 0.5\,nA).  
(b). Top view of the DFT-LDA slab consisting of six Si-layers
    sandwiched between one top Sn-layer and one bottom H-layer (not
    shown in this figure). Sn atoms align in a plane.  
    A 9-site Sn cluster is used in the DCA calculations (see text).}
\label{Fig1}
\end{figure}

\begin{figure}[htbp]
\centering
\includegraphics[width=\linewidth]{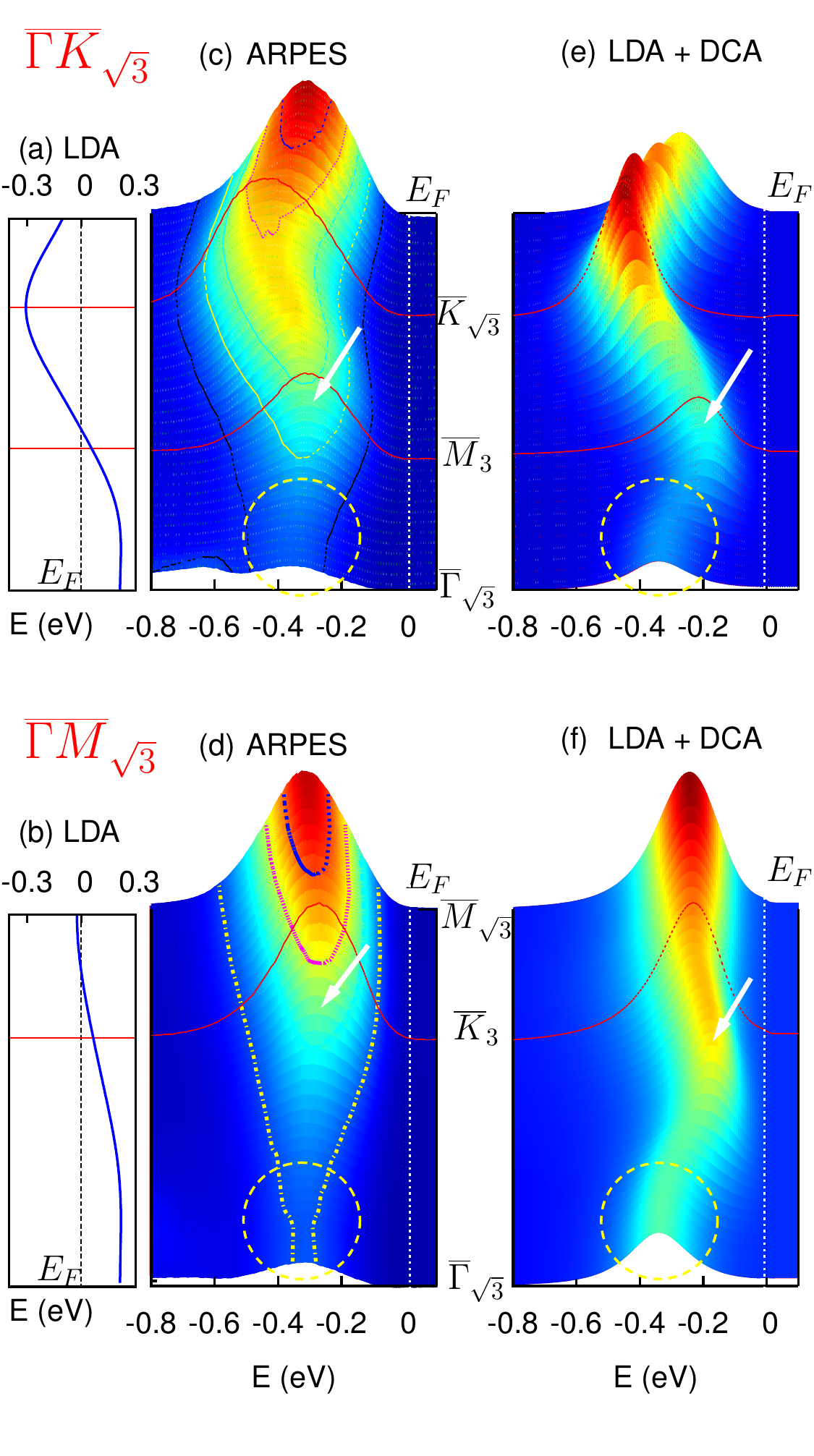}
\caption{(a), (b) LDA-derived band structure along the two major high symmetry
  directions $\overline{\Gamma K}_{\sqrt{3}}$ and $\overline{\Gamma M}_{\sqrt{3}}$ 
  in comparison with (c), (d) corresponding photoemission spectra ($T
  \approx 60$\,K, $h\nu=130$\,eV) and (e), (f) the spectral
  function as obtained from LDA + DCA calculations. In (c) and (d),
  the contour lines are shown as a guide to the eye. White arrows in
  (c - f) indicate additional band maxima, which are not present in
  the LDA-derived band at the same energies. The four yellow circles around 
  $\overline{\Gamma}_{\sqrt{3}}$ in (c - f) highlight the shadow bands observed
  in both ARPES and in the LDA + DCA calculations, which are also missing
  in the LDA-derived band.} 
\label{EDC_Comp}
\end{figure}

Fig. \ref{Fig1} shows our experimental and
theoretical situation for the study of the Sn/Si(111) system.   
Experimentally, clean surfaces were obtained by thermal desorption of
the capping oxide at $ \approx 
1250\,^{\circ}\mathrm{C}$ from n-type Si substrates ($\rho<
0.01\,\Omega \mathrm{cm}$) resulting in a sharp $7 \times 7$ pattern
in LEED.  
Subsequently, 1/3 monolayer Sn was deposited on the
substrates by electron beam evaporation. 
After an anneal at $\approx 700\,^{\circ}\mathrm{C}$, the
Sn/Si(111)-$\sqrt{3}\times\sqrt{3}$
surface reconstruction was verified in high quality by LEED and STM,
see Fig. \ref{Fig1}(a). 
Theoretically, we construct a slab with six Si layers and saturate the Si
bottom layer by hydrogen atoms.  
Sn atoms are placed on the top layer at equivalent $T_{4}$ lattice
sites, see Fig. \ref{Fig1}(b). This is in agreement with the
atomic structure derived from surface x-ray diffraction
\cite{Conway1989}. 
Therefore, we exclude any additional lattice superstructure, e.g., a
$3 \times 3$ periodicity, in our calculations and
expose our system to the electron-electron interaction $U$ only. 

First, we carry out {\it ab initio} DFT
calculations based on LDA
\cite{PhysRevB.83.041104}, which predict a half-filled metallic ground state
of Sn/Si(111), see Fig. \ref{EDC_Comp}(a) and (b). 
This is well in agreement with preceding LDA calculations
\cite{PhysRevB.59.1891, PhysRevB.62.1556, PhysRevB.82.035116}.
Next, we project the Sn-related surface band onto the maximally
localized Wannier basis and construct a single-band Hubbard model. 
The dynamical cluster approximation
(DCA)~\cite{RevModPhys.77.1027} is used with the
continuous-time quantum Monte Carlo method \cite{PhysRevB.72.035122}
to solve this model in the paramagnetic phase.  
We consider a 9-site cluster in our calculations, as shown in
Fig. \ref{Fig1}(b). 
The surface Brillouin zone (SBZ) is divided into 9 equal-area sectors.
In each of them the electron self-energy becomes a constant,
i.e., $\Sigma(k, i\omega_{n})\rightarrow\Sigma(K, i\omega_{n})$.         
Our LDA + DCA calculations reported here are the first to examine
the spectral weight redistribution in momentum space in the
thermodynamic limit for Sn/Si(111).
In what follows, we address the theoretical photoemission spectra,
which are related in the usual way to the imaginary part of the
single-particle Green's function. 
The latter is calculated following a recipe contained in
Ref.~\cite{0295-5075-85-5-57009}.

\begin{figure*}
\centering
\includegraphics[width=0.38\linewidth]{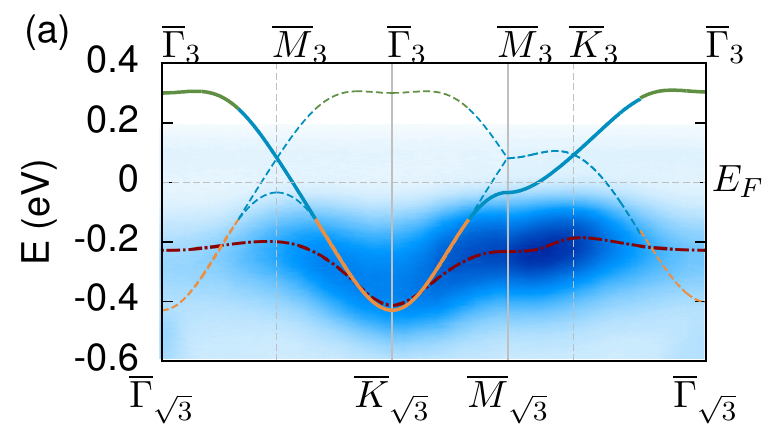}%
\includegraphics[width=0.24\linewidth]{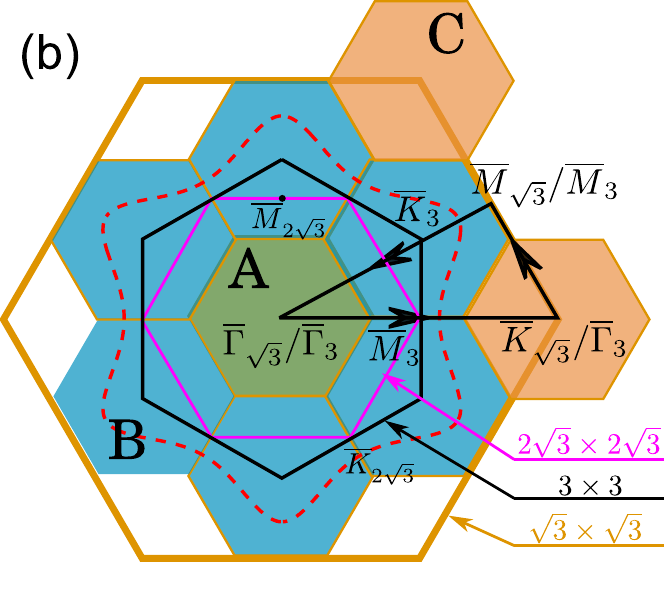}%
\includegraphics[width=0.38\linewidth]{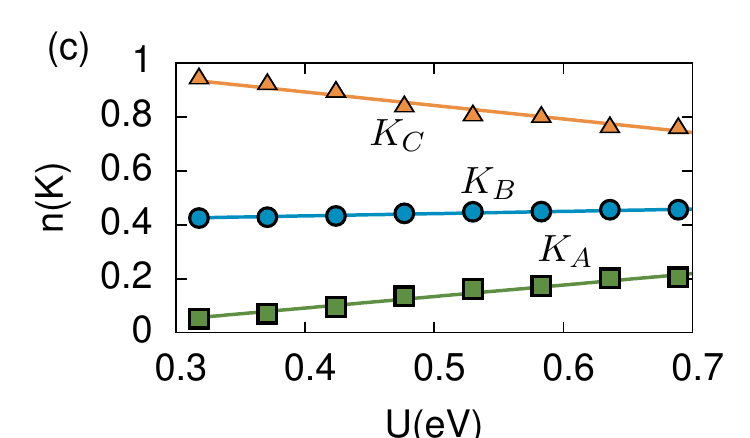}
\caption{ 
  (a). LDA band structure (solid lines),
  backfolded LDA bands (dashed lines) with respect to $3 \times 3$ SBZ
  boundaries, and ARPES band map (false color plot) with intensity maximum
  (red dash-dotted line) indicated as guide to the eye. 
  In order to achieve a better agreement with the experimental band edge
  at $\overline{K}_{\sqrt{3}}$, we have to scale the LDA-band, which is shown as
  a solid line in Fig.~\ref{BackFolding}(a), by a factor of 1.3. 
  This scaling effect has essentially been captured by the LDA + DCA
  calculations in Fig.~\ref{EDC_Comp}.  
  This factor should not be interpreted as the effective mass of this
  surface band, since Fermi liquid theory fails to explain the
  shadow band.
  (b). The 1st $\sqrt{3} \times \sqrt{3}$ SBZ of the Sn/Si(111)
  surface, which is divided into three different sectors (A, B, C) in
  the DCA approximation, see text for more details.
  The black and violet hexagons correspond to the SBZ of a $3 \times
  3$ and $2\sqrt{3}\times2\sqrt{3}$ superstructure. 
  (c). Average particle density in each momentum sector as a
  function of interaction, reflecting the weight transfer from sector
  $K_C$ to $K_A$.
  }  
\label{BackFolding}
\end{figure*}
Fig. \ref{EDC_Comp} contains the experimental and theoretical
photoemission spectra as a function of position in the SBZ. 
ARPES experiments were carried out at the electronic structure 
factory (ESF) endstation of beamline 7.0.1 at the Advanced Light
Source (ALS), which provides sample cooling down to 10\,K
and is equipped with a 6-axis goniometer. Photoelectrons were detected
with a Scienta R4000 spherical analyzer with energy resolution set to
25\,meV throughout all measurements. 
In the theoretical spectra calculations, we set
$U\sim0.66$\,eV, which is slightly above $U_{c} \approx 0.60$\,eV for the
metal-insulator transition (MIT) in this system \cite{PhysRevB.83.041104}. 
In a related system, i.e. Sn/Ge(111), earlier studies reported two
distinct surface states in ARPES \cite{PhysRevLett.81.2108,
  PhysRevLett.82.442}.
Sn atoms were believed to stay at two different adsorption sites
\cite{PhysRevLett.86.4891, PhysRevLett.96.126103,
  PhysRevLett.101.186102} in this system.
However, in the Sn/Si(111) surface band, such a splitting is very small  
at low temperature and even absent at room temperature \cite{PhysRevB.62.8082,
  PhysRevB.68.235332}.  
We do not want to study such small energy differences in our
theoretical spectral function, which contains the uncertainty from the
analytical continuation. 
However, it should be noted that quantum Monte Carlo calculations have
shown that strong electron repulsions can further
split the lower Hubbard band \cite{PhysRevB.62.4336, 
  PhysRevLett.79.1122}. 
Thus, whether or not the two peak structure in that system should be
attributed to a geometrical change deserves a more careful study.  
Moreover, it was shown in a recent theoretical work that electron
correlations alone can induce structural transformations in elemental
iron \cite{2011arXiv1110.0439L}. 
Here, we will not address such additional effects. For Sn/Si(111) we
compare ARPES and theory by focusing on the band topology and its
k-dependent spectral weight.  

Two ARPES k-space line scans along $\overline{\Gamma K}_{\sqrt{3}}$
and $\overline{\Gamma M}_{\sqrt{3}}$ are shown next to the corresponding LDA + DCA results in
Fig. \ref{EDC_Comp}(c) to (f). Rather good overall agreement can be detected in this
comparison. Along both $\overline{\Gamma K}_{\sqrt{3}}$ and $\overline{\Gamma
  M}_{\sqrt{3}}$, we observe a shadow band in ARPES
  around $\overline{\Gamma}_{\sqrt{3}}$, as indicated by yellow
circles in Fig. \ref{EDC_Comp}(c) and (d).
The evolution of this band clearly shows an \emph{additional band maximum}  
at a position close to $\overline{M}_{3}$ along
$\overline{\Gamma K}_{\sqrt{3}}$ and in vicinity of $\overline{K}_{3}$ along
$\overline{\Gamma M}_{\sqrt{3}}$ indicated by the white arrows.
$\overline{M}_{3}$ and $\overline{K}_{3}$ are high symmetry points of
a $3\times3$ SBZ. 
The appearance of the additional band maximum, which is absent in the
LDA results in Fig. \ref{EDC_Comp}(a) and (b), modifies the
spectral dispersion from  $\sqrt{3}\times\sqrt{3}$ to an approximate
$3\times3$ symmetry.   
In the LDA + DCA calculations shown in Fig.~\ref{EDC_Comp}(e) and (f),
both the \emph{shadow band} and its spectral evolution as a function
of momentum are well reproduced. 
The shadow band is clearly visible in theory and slightly more
pronounced than in the experimental spectra.
The overall agreement between ARPES and the LDA + DCA calculations,
and especially the appearance of this shadow band, represent strong 
evidence of many-body effects in this system. In contrast, a structural
origin is rather unlikely since in our calculations all Sn atoms are
located at equivalent lattice sites within the same atomic layer.   
Thus, the approximate $3\times3$ symmetry cannot result from any
structural change, which is essentially in accordance with the already
mentioned absence of a surface-band splitting in the Sn/Si(111) system
at low temperatures.  

Profeta and Tosatti suggested that the $120^{\circ}$-AFM ordering and consequent
folding might be the origin of the shadow band \cite{PhysRevLett.98.086401}. 
In what follows, we want to demonstrate that it is very likely that
this system is magnetically short range ordered. However, a strict
$120^{\circ}$-AFM order cannot fully explain the surface band topology,
especially the energy range of the shadow band and the position of
the band maximum which is not located exactly at $\overline{K}_{3}$ along the
$\overline{\Gamma M}_{\sqrt{3}}$ direction. 
According to Profeta and Tosatti, we  
backfold the original LDA-band into the 1st SBZ with respect to the
$3 \times 3$ SBZ boundary, which then corresponds to a $120\,^{\circ}$-AFM order.    
In Fig. \ref{BackFolding}(a), the original LDA-band is plotted as a
solid line along the $\overline{\Gamma KM\Gamma}_{\sqrt{3}}$
directions, the dashed lines are the folded bands. 
The high symmetry points of the $3 \times 3$ SBZ are labeled as
$\overline{\Gamma}_{3}, \overline{M}_{3}, \overline{K}_{3}$. 
Three different colors on the LDA and the backfolded bands are used to
indicate different momentum sectors, as those in
Fig. \ref{BackFolding}(b).   
These three inequivalent momentum sectors are derived from the symmetry
preserved in the 9-site cluster DCA calculations, which 
are labeled as sector A, B and C. 
The high symmetry points, i.e., $\overline{\Gamma}_{\sqrt{3}},
\overline{M}_{\sqrt{3}}$ and $\overline{K}_{\sqrt{3}}$, are contained
in sector A, B and C, respectively.   
The red dashed line represents the Fermi surface (FS) of the LDA band, which
is completely contained in sector B.      
Evidently, in the $3 \times 3$
SBZ a band is located around $\overline{\Gamma}_{\sqrt{3}}$ at $E < 0$,
which is obviously back-folded from $\overline{K}_{\sqrt{3}}$,
see Fig. \ref{BackFolding}(a).

First of all, our calculations strongly support the existence of a
magnetic order and the consequent band back-folding.
In Fig. \ref{BackFolding}(c), the average particle numbers $n(k)$ at each
momentum sector are shown as a function of the interaction $U$.  
$n(k)$ was directly calculated in the LDA + DCA.
It relates to the spectral function $A(k, E)$ by $n(k) =
\int_{-\infty}^{\infty}\mathrm{d} E A(k, E)f(E, T)$,
where $E$ denotes the energy and $f(E, T)$ the Fermi function.
We found qualitatively a different behavior of $n(k)$, when $k$~$\in$~$K_{A},
K_{B}$ or $K_{C}$. 
$n(K_{A})$ monotonically grows with the increase of the
interaction, while $n(K_C)$ behaves exactly the opposite.  
In contrast, $n(K_B)$ stays almost constant while varying $U$.
In sector B, with the increase of $U$, the quasiparticle peak at the
Fermi level gradually looses its weight until a \emph{charge gap}
opens.  
The constant value of $n(K_{B})$, therefore, strongly indicates that
the total spectral weight in sector B does not change, however, the
spectral weight lost at the quasiparticle peak transfers to the lower
and upper Hubbard bands within this sector.   
The constant value of $n(K_{B})$ is a strong indication of the
\emph{Mott type} of the MIT.   
Moreover, an increasing $U$ results in a spectral weight
transfer from sector C to sector A.  
In the $U=0$ limit, there is no intensity at
$\overline{\Gamma}_{\sqrt{3}}$ for energies $E < 0$, giving the
almost zero value of $n(K_{A})$. 
For higher values of $U$, part of the spectral weight around
$\overline{K}_{\sqrt{3}}$ transfers to $\overline{\Gamma}_{\sqrt{3}}$,  
resulting in the increasing/decreasing behaviors of
$n(K_{A})$/$n(K_{C})$. 
Thus, what we observe from Fig. \ref{BackFolding}(b) mainly
reflects the spectral weight transfer from $\overline{K}_{\sqrt{3}}$
to $\overline{\Gamma}_{\sqrt{3}}$, which strongly supports the
\emph{band back-folding} picture. 

However, a strict $3\times3$ symmetry cannot fully explain the shadow
band we observed.
A close comparison of ARPES and the folded LDA bands reveals
that the shadow band at $\overline{\Gamma}_{\sqrt{3}}$ is not at the
same energy as the band at $\overline{K}_{\sqrt{3}}$, as it would be
if the magnetic order was $120\,^{\circ}$-AFM. 
The shadow band stays at higher energies than that at
$\overline{K}_{\sqrt{3}}$.
Thus, the magnetic SBZ of Sn/Si(111)-$\sqrt{3}\times\sqrt{3}R30^{\circ}$ can
only be approximate to $3\times3$. 
It reflects that the magnetic order derived from our ARPES and
calculations is actually close to but different from the classic
$120^{\circ}$-AFM for triangular systems.   
This is partially due to additional hopping processes inherently
contained in Sn/Si(111) as compared to the ideal triangular model 
with nearest-neighbor hopping only.
In our previous study, we found that the inclusion of the
next-nearest-neighbor hopping in triangular lattices changes the
spin-susceptibility peak-position from near $\overline{K}_{\sqrt{3}}$
to $\overline{M}_{\sqrt{3}}$ in the 1st SBZ which indicates a magnetic
order change from $120^{\circ}$-AFM to a \emph{row-wise} (RW) type AFM 
~\cite{PhysRevB.83.041104}.
This is equivalent to the magnetic ordering in the Mn/Cu(111) surface, which can
be effectively described by a triangular Heisenberg model with higher
order exchange interactions \cite{PhysRevLett.86.1106}.
The superposition of three equivalent spin arrangements of RW-AFM order,
which possess a $2\sqrt{3}\times 2\sqrt{3}$ magnetic SBZ , can
further lower the total energy and is thus favored.
$\overline{M}_{3}$ of the $3\times3$ SBZ is also a high symmetry point
of the $2\sqrt{3}\times 2\sqrt{3}$ SBZ. However, as can be seen from
Fig. \ref{BackFolding} (b), $\overline{M}_{2\sqrt{3}}$ of
$2\sqrt{3}\times 2\sqrt{3}$ is close to but different from
$\overline{K}_{3}$ of the $3\times3$ SBZ, 
which essentially explains the agreement of the band maximum position
with $\overline{M}_{3}$ along $\overline{\Gamma M}_{\sqrt{3}}$ and the
discrepancy with $\overline{K}_{3}$ along $\overline{\Gamma K}_{3}$.
The current surface-band topology study and the spin susceptibility
calculations in our previous work~\cite{PhysRevB.83.041104} coincide
with each other, and both point at the magnetic order to be of RW-AFM type.
To this end, a spin symmetry-broken many-body calculation and a
spin-resolved STM study are highly desirable for further understanding of
this adatom system.   

\begin{figure}[htbp]
\centering
\includegraphics[width=0.9\linewidth]{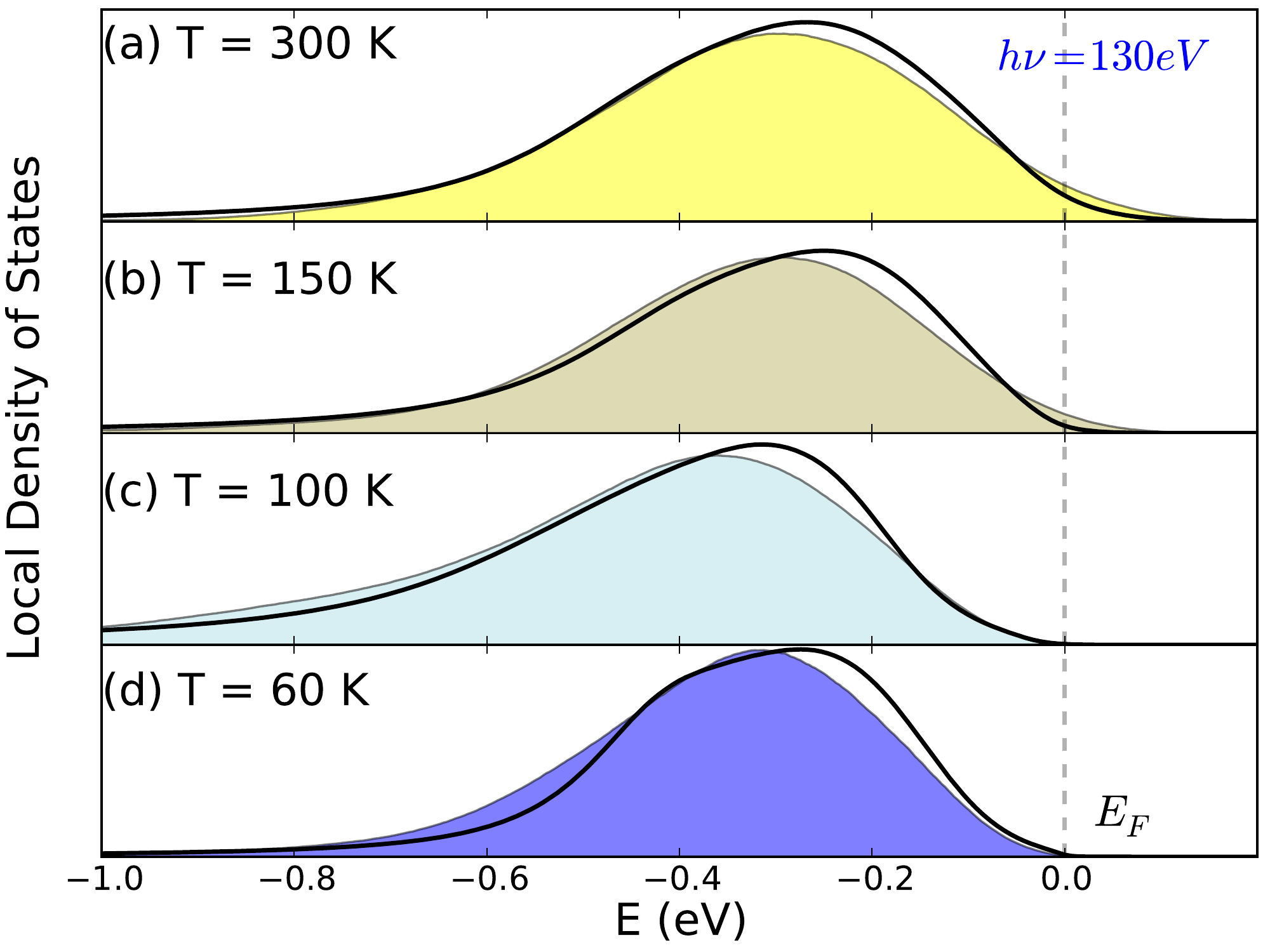}
\caption{Comparison of the angle-integrated ARPES spectra
  (color filled area) with the local density of states 
  calculated from the LDA + DCA at $U = 0.66$\,eV (solid black lines)
  for different temperatures. }  
\label{Fig:LDOS}
\end{figure}
In Fig. \ref{Fig:LDOS}, we show a comparison of the
  temperature dependent angle-integrated photoemission spectra with
  the local density of states (LDOS) calculated from the LDA + DCA. 
  The experimental curves are obtained by integrating the ARPES
results along the 
$\overline{\Gamma}_{\sqrt{3}}\rightarrow
\overline{M}_{\sqrt{3}}\rightarrow\overline{K}_{\sqrt{3}}
\rightarrow\overline{\Gamma}_{\sqrt{3}}$
directions.  
Experimentally, we observe a MIT between $T=100$\,K and 150\,K.  
The angle-integrated spectral weight at the Fermi level becomes
nonzero with increasing temperature. 
This is in agreement with the many-body description of a Mott MIT.
On the other hand, higher temperature has almost no effect on the
spectra far away from $E_F$. 
We find a marginal change on the spectra for
energies below $-0.6$\,eV. In contrast, for increasing temperatures
spectral weight is transferred towards the Fermi level, driving this
surface system from insulator to metal.     
Theoretically, for $U  = 0.66$\,eV we obtain an overall good agreement
with the experiments for  
different temperatures. At all temperatures represented in
Fig. \ref{Fig:LDOS}, the part of the LDOS close to the Fermi level
coincides well with its experimental counterpart.   

In summary, we have shown that the key features of
a triangular adatom system, realized by the
Sn/Si(111)-$\sqrt{3}\times\sqrt{3}R30^{\circ}$ surface, can be
qualitatively explained by strong electronic correlations.   
By assuming a planar configuration of the Sn atoms, we find a good
overall agreement between experiment and the LDA + DCA
calculations. 
A temperature dependent MIT is found, which closely coincides with the
Mott description for this surface system.  
We find strong evidence for a spectral weight transfer from the
momentum region around $\overline{K}_{\sqrt{3}}$ to 
$\overline{\Gamma}_{\sqrt{3}}$, indicating the existence of 
a magnetic order in this system.
The additional symmetry observed in ARPES can then be understood
as a band back-folding with respect to a new magnetic ordering.
This should stimulate further studies on the magnetic
properties of this and related systems. 

This work is financially supported by the Deutsche
Forschungsgemeinschaft under grant FOR 1162.

\bibliographystyle{apsrev4-1}	
\bibliography{ref}

\end{document}